\documentclass{CSML}

\def\dOi{12(1:3)2016}
\lmcsheading%
{\dOi}
{1--23}
{}
{}
{Jun.~24, 2014}
{Mar.~\phantom09, 2016}
{}

\ACMCCS{[{\bf Theory of computation}]: Semantics and
  reasoning---Program semantics---Denotational Semantics;
  Logic---Linear logic}

\usepackage{etex}
\usepackage{todonotes}
\usepackage[english]{babel}
\usepackage[utf8]{inputenc}
\usepackage{xspace}
\usepackage{amsfonts}
\usepackage{amsmath}
\usepackage{amssymb}
\usepackage{MnSymbol}
\usepackage{aeguill}
\usepackage{hyperref}
\usepackage{amsthm}
\usepackage{stmaryrd}

\usepackage{float}
\floatstyle{boxed} 
\restylefloat{figure}

\input{diagxy}
\usepackage{bussproofs}
\usepackage{cmll}

\newcommand{\lct}{{lctvs}\xspace}
\newcommand{\wstar}{{weak*}\xspace}

\newcommand{\lin}{\mathcal{L}}
\newcommand{\blin}{\mathcal{B}}

\newcommand{\nlin}{\mathcal{L}^n}
\newcommand{\plin}{\mathcal{L}^p}
\newcommand{\mlin}{\mathcal{L}^m}
\newcommand{\nmon}{\mathcal{H}^n}
\newcommand{\pmon}{\mathcal{H}^p}
\newcommand{\mmon}{\mathcal{H}^m}

\newcommand{\jmon}{\mathcal{H}^j}

\newcommand{\CC}{\mathbb{C}}
\newcommand{\W}{\mathcal{W}}
\newcommand{\KK}{\mathbb{K}}
\newcommand{\NN}{\mathbb{N}}
\newcommand{\Weak}{\mathbb{W}\mathrm{eak}}
\newcommand{\Ker}{\mathrm{Ker}}

\newcommand{\shiftup}{\mathord{\uparrow}}
\newcommand{\shiftdown}{\mathord{\downarrow}}

\begin{document}

\author[M.~Kerjean]{Marie Kerjean} 
\address{Univ Paris Diderot, Sorbonne Paris Cité, PPS, UMR 7126, CNRS, F-75205 Paris, France}
\email{marie.kerjean@pps.univ-paris-diderot.fr}

\title[Weak topologies for linear logic]
      {Weak topologies for linear logic}


\begin{abstract}
\noindent
We construct a denotational model of linear logic, whose objects are all the locally convex and separated topological vector spaces endowed with their weak topology. The negation is interpreted as the dual, linear proofs are interpreted as continuous linear functions, and non-linear proofs as sequences of monomials. We do not complete our constructions by a double-orthogonality operation. This yields an interpretation of the polarity of the connectives in terms of topology. 
\end{abstract}

\maketitle

\section*{Introduction}

Linear logic can be seen as a refinement of intuitionistic logic, providing an analysis of classical logic through the notion of polarities and involutive linear negation \cite{LR03}. The linearity hypothesis was made by Girard \cite{Gir87} after an investigation of models of lambda-calculus \cite{Gir88}. Models of linear logic have provided a fresh point of view and new intuitions that was applied to traditional fields of study, such as game semantics \cite{AM99}. It even led to the discovery of new computational languages, such as differential $\lambda$-calculus \cite{ER03}.

One of the challenges in the semantical study of linear logic is to get closer to the algebraic intuitions of the syntax. This is done by interpreting (linear) proofs as linear maps between vector spaces \cite{Blu96,Ehr02,Gir04,Ehr05,BET12}. There is one standard issue with this kind of work : the interpretation of the exponential connectives asks for infinite dimensional spaces, while interpreting the involutive linear negation is much easier in finite dimensional spaces. Indeed, the equivalence between a formula and its double negation in linear logic asks for the vector spaces considered to be isomorphic to their double dual. This is immediate when the spaces are finite dimensional, but much harder to obtain for infinite dimensional vector spaces. 

Models of linear logic are often inspired by coherent spaces, or by the relational model of linear logic. Coherent Banach spaces \cite{Gir96}, coherent probabilistic or coherent quantum spaces \cite{Gir04} are Girard's attempts to extend the first model, as finitenes spaces \cite{Ehr05} or K\" othe spaces \cite{Ehr02} were designed by Ehrhard as a vectorial version of the relational model. 
 
Our approach takes another direction, and provides a model capturing a very large class of vector spaces. We propose to interpret formulas, which are isomorphic to their double negation, by any locally convex Hausdorff \emph{topological} vector space, endowed with a specific topology. It allow us to step away from the combinatorial point of view, and introduce a basis-free vectorial interpretation of the connectives of linear logic. As in Scott domains, we interpret proofs by continuous functions. More precisely, proofs will be interpreted by linear continuous functions between topological vector spaces. We interpret the linear negation of a formula by the space of all continuous linear forms on the interpretation of the formula (that is, the topological dual of this space). 

We do not settle for a model of linear logic (LL) obtained by a Chu construction \cite{Bar79}. We want the involutivity of negation to be an intrinsic property of our objects. This leads us to the only real restricting choice of this paper: we endow our spaces with their weak topology. The weak topology on a topological vector space $E$ is the coarsest topology induced by its dual : a sequence of elements of $E$ converges for the weak topology if and only if it converges when composed with any (continuous) linear form on $E$. A symmetric definition yields the weak* topology on $E'$. Endowing a space with its weak topology and its dual with its weak* topology result in an isomorphism between the space and its double dual. Every other construction in the interpretation of LL will stem from this first choice.

Note that we don't ask for our spaces to be Cauchy-complete, thus there is no completion on the interpretation of our connectives. This reduces drastically the possibilities for the interpretation of the exponential. Indeed, the latter is tied to the induced (co-Kleisli) cartesian closed category of non-linear functions. To deal with these non-linear functions, completion is usually necessary \cite{Ehr05,BET12,KT15}. The lack of completion explains the form of our interpretation of non-linear proofs, as simple sequences of monomials. Thus, even though the model interprets differential linear logic, we are far from capturing the intuition of smoothness of proofs behind the notion of differentiation (as done for example in \cite{BET12,KT15}).

Most models of linear logic (denotational semantics \cite{Gir87,Ehr02, Ehr05}, phase semantics \cite{Gir87}, ludics \cite{Gir01, Ter11}, Geometry of Interaction \cite{Gir11, Sei12}) interpret negation through an orthogonality relation. The negation of a space $E$ is the set of all elements which are in the orthogonality relation with all elements of $E$. This is not the way the duality is constructed here: reflexivity is not based on an orthogonality. 
Polarity of the connectives matters nonetheless when endowing a space with its weak topology, as when interpreting the negation as an orthogonality. The negative connectives are those who preserve the fact to be endowed with the weak topology, while the positive connectives are those who need to be applied a shift, that is a change of polarity corresponding in the model to the enforcement of weak topology.

\subsubsection*{Synthesis of the constructions}
Our constructions are very simple, as they only use well-known tools of the theory of topological vector spaces. Formulas of linear logic are interpreted by any locally convex and separated topological vector space, endowed with its weak topology. The negation of a formula is interpreted by the dual of the interpretation of this formula, endowed with its \wstar topology.  

The multiplicative conjunction $\otimes$ is interpreted by a specific topological tensor product endowed with its weak topology: choosing the topology of the algebraic tensor product is indeed one of the determining steps in the construction of this model. The $\parr$ is interpreted as the topological dual of $\otimes$. As a result of these constructions, the type of linear proofs between two formulas is interpreted as the space of linear continuous functions between the interpretation of these formulas, endowed with the topology of simple convergence.

As for additive connectives, $\with$ is interpreted by the binary topological product, and $\oplus$ by the binary topological co-product endowed once again with its weak topology. They coincide on finite indexes.

Finally, the exponential is constructed so that non-linear proofs between two spaces are interpreted by the sequences of monomials between these two spaces. This construction follows the idea of quantitative semantics, which is at the heart of linear logic \cite{Gir88}.

\subsubsection*{Related works}

The linear negation is often interpreted with an orthogonality relation \cite{Ehr02,Ehr05,Gir04} or with a Chu construction as in coherent Banach spaces \cite{Gir96}. 

The construction presented here is very general, as any locally convex and separated topological vector space is turned into an object of our category. Our approach differs from the one presented in the finiteness spaces \cite{Ehr05}, or in the Hopf algebras as model of multiplicative linear logic \cite{Blu96}: the topologies used there are Lefschetz topologies, that is topologies where neighbourhoods of $0$ are sub-vector spaces, opposed to the intuitive idea of unit ball coming from normed spaces.   

This generality allows us to define our tensor product as an algebraic tensor product, and not as its bidual or biorthogonal, contrasting with what happens in finiteness spaces \cite{Ehr02} or K\" othe spaces \cite{Ehr05}. On the contrary to what happens in Ehrhard's model, our constructions are basis-free. Note that K\"othe spaces \cite{Ehr02} are endowed with a polar topology (the normal topology) which is in general finer than the weak topology.

Moreover, the interpretation of the classical duality is internalized, and not obtained as the result of a Chu construction as in coherent Banach spaces \cite{Gir96} or as in works by M. Barr \cite{Bar00, Bar79, Bar91}. With an adjunction between Chu spaces and the category of topological vector spaces, Barr obtains a $\ast$-autonomous category of spaces endowed with their weak topology, where the spaces of linear functions are the same as ours. However, the tensor product is completed, as $ E \otimes F $ is defined as $  \lin(E,F')'$. Here, our tensor product is not completed, as $E \otimes F$ is the algebraic tensor product endowed with some topology, and our constructions avoid the digression through the Chu category. This work can therefore be seen as an extension of Barr's work to a Seely category.

One could think of the interpretation of polarized linear logic (LLP) in a control-category by its negative connectives, described by O. Laurent in his thesis \cite{Lau02}. However, this is not what is used here, as positive connectives are not primarily interpreted as the dual of the interpretation of their negation. This model neither corresponds to the interpretation of LLP in a co-control category, as positive connectives do not preserve the property of being endowed with one's weak topology.

\tableofcontents

\section{Topologies on vector spaces and spaces of functions}

Recall that a topological vector space is a vector space endowed with a topology making the sum and multiplication by a scalar continuous \cite[I.2.1]{Jar81}. The vector space is said to be Hausdorff when the topology is so, that is when given any two distinct point in the vector space, they belong respectively to two open sets with empty intersection. 

In a topological space, a basis $\mathcal{U}$ of open sets is a collection of open sets such that every open set is the union of elements of $\mathcal{U}$. In a topological vector space, we only need to know the neighbourhoods of $0$ to retrieve the entire topology, as the addition is continuous. Therefore, we will most of the time describe the topology on our spaces by giving a basis of $0$-neighbourhoods, or some time by giving a subbasis of $0$-neighbourhoods. A subbasis $\mathcal{U}$ of a vector topology is a collection of open sets such that every open set is the union of finite intersections of elements of $\mathcal{U}$. See the definition of the \wstar topology for an example of basis and subbasis of an topological vector space. \newline

\begin{defi}{\protect{\cite[6.7]{Jar81}}}
A Hausdorff locally convex topological vector space is a Hausdorff topological vector space with a subbasis of neighbourhoods of $0$ consisting of convex subsets. We write \lct to denote such vector spaces.
\end{defi}

From now on, $E$, $F$ and $G$ are \lct. All our \lct are vector spaces over $\mathbb{K}=\mathbb{R}$ or $\mathbb{K}=\mathbb{C}$.

Let us begin with the heart of our construction, that is the dual $E'$ of a vector space $E$. The space $E'$ is fundamental, as it defines the weak topology on $E$, and as the topology on itself allows us to interpret the classical duality. 

\begin{defi}
\label{defi:weaktopo}
If $E$ is a \lct, we will denote by $E'$ the space of all continuous linear forms $ l: E \rightarrow \mathbb{K}$ on $E$, endowed with its \wstar topology. The space $E'$ is called the dual of $E$, or the topological dual of $E$ when there is an ambiguity. 
\end{defi}

\subsection{Weak and \wstar topologies}
We will endow our spaces with their weak topology, that is with the topology generated by their continuous duals. The original topology of a space $E$ is also called its strong topology, and it allows to compute the continuous dual $E'$. Practically all the definitions presented here are well-known definition from functional analysis, and proofs can be found for example in the literature \cite{Jar81,Kot69}.

\begin{defi}
The \wstar topology on $E'$ is the topology of pointwise convergence on $E$. The \wstar topology is also the inductive topology generated by $E$ on $E'$, that is the coarsest topology on $E$' making the evaluation functions $ l \in E'  \mapsto l(x)$ continuous for all $x \in E$. 
\end{defi}
A basis for the \wstar topology on $E'$ is the collection of all $$\mathcal{W}_{x_1,...x_n, \epsilon} = \left\lbrace l \in E' ~|~~|l(x_1)| < \epsilon ,...|l(x_n)| < \epsilon \right\rbrace$$
 where $n \in \mathbb{N}$, $x_i \in E$ and $\epsilon >0$.
A subbasis for the \wstar topology is the collection of all 
$$W_{x, \epsilon} = \left\lbrace l \in E' ~|~~|l(x)| < \epsilon  \right\rbrace. $$

\begin{defi}
\label{weak_topo}
The weak topology on a \lct $E$ is the inductive topology generated by $E'$, that is the coarsest topology on $E$ making all the functions $l \in E'$ continuous.
\end{defi}

A basis for the weak topology on $E$ is the collection of all $$\W_{l_1,...l_n, \epsilon} =  \left\lbrace x \in E ~|~~|l_1(x)| < \epsilon ,...|l_n(x)| < \epsilon \right\rbrace$$
 where $n \in \mathbb{N}$, $l_i \in E'$ and $\epsilon >0$. Let us generalize the description of the basis of the weak and \wstar topology.

Consider $\mathcal{F}(E,F)$ a vector space of functions between $E$ and $F$. When $B \subset E$ and $U \subset F$, we write 
$$\W_{B,U} = \left\lbrace f \in \mathcal{F}(E,F) ~|~ f(B) \subset U \right\rbrace .  $$ 
When $B = \{x_1,...,x_n \}$ is finite, then  $\W_{B,U}$ is written $\W_{x_1,..x_n,U}$. When $F = \KK$ and $U = \{ y \in F ~|~~|y| < \epsilon \}$, then $\W_{B,U}$ is written $\W_{B,\epsilon}$.
Note that algebraically $E$ can be considered as a sub-vector space of $E''$ through the application $ ev: x \mapsto ( ev_x: l \mapsto l(x))$. The notation used in the definition \ref{defi:weaktopo} is then coherent with the one described above. 
Note that $\W_{B,U}$ is convex as soon as $U$ is convex.

\begin{nota}
From now on $E_w$ will denote a \lct endowed with its weak topology. The original topology on $E$ is called its strong topology.
\end{nota}
 
\begin{prop}
When $E$ is a \lct, then so are $E_w$ and $E'$.
\end{prop}

\proof
As a consequence of Hahn-Banach separation theorem \cite[7.2.2.a]{Jar81}, we have that $E'$ separates the points of $E$: if $x,y  \in E$ are distinct, then there is $l \in E'$ such that $l(x) \neq l(y)$. This makes $E$ endowed with its weak topology a Hausdorff topological vector space. It is locally convex as $\W_{B,U}$ is convex as soon as $U$ is convex.
The same arguments make $E'$ a \lct.
\qed

\begin{nota}
We work in the category of \lct endowed with their weak topology and continuous linear maps. Let us denote this category by $\Weak$. The relation $\simeq$ denotes an isomorphism in $\Weak$ between two \lct. When we need to speak about an isomorphism in the category of vector spaces and linear maps, we will use the symbol $\sim$.
\end{nota}

In $\Weak$ the spaces $(E_w)'$ and $E'$ corresponds.

\begin{prop}{\cite[II.8.1.2]{Jar81}}
\label{w_wstar}
For any \lct E, $(E_w)'$ is linearly homeomorphic to $E'$. That is, $(E_w)' \simeq E'$. \qed
\end{prop}

The demonstration of this proposition uses the following lemma:

\begin{lem}
\label{lem_alg}
Consider $E$ a vector space and $l,l_1, ... l_n$ linear forms on $E$. Then $l$ is in the vector spaces generated by the family $l_1, \dots, l_n$ (denoted $\mathrm{Vect} (l_1,...l_n)$) if and only if $\bigcap_{k=1}^n \Ker(l_k) \subset \Ker(l)$. \qed
\end{lem}
We can continue to write $E'$ for the dual of a space $E$, regardless whether it may be endowed with its weak topology. We will write $E'$ for $(E_w)'$ and $E'_w$ for $(E')_w$.

\subsection{Reflexivity}

Let us define 

\begin{equation*}
ev: 
\left\lbrace
\begin{split}
E & \rightarrow E'' \\
x & \mapsto ev_x = (l \in E' \mapsto l(x))
\end{split}
\right.
\end{equation*}

The function $ev$ is linear and injective as $E'$ separates $E$. It is continuous and open as both $E$ (resp $E''$) are endowed with the weak (resp. \wstar) topology induced by $E'$. \newline

The starting point of this paper is the fact that when $E'$ is endowed with its \wstar topology, $E$ can be considered as a reflexive space, that is $E_w \simeq E''_w $. This equality models the involutive linear negation of linear logic, and will make our category of topological vector spaces and linear maps $\ast$-autonomous.

\begin{prop}{\cite[II.8.1.2]{Jar81}}
The function $ev$ is an isomorphism from $E$ to $E''$.
\end{prop}

\proof
The proof is done as in the Proposition \ref{w_wstar}, using Lemma \ref{lem_alg}. The key to this proof is the fact that $E'$ is endowed with the \wstar topology, and thus for every $l \in E''$ there is $x_1, ... , x_n \in E$ such that 
\[l( \mathcal{W}_{x_1,...x_n, 1}) \subset \{ x \in \KK ~|~|x|<1 \}\, .\]
This implies $\bigcap_i \Ker(ev_{x_i}) \subset \Ker (l)$, and thus through Lemma \ref{lem_alg} $ l \in \mathrm{Vect}(ev_{x_i})$. That is, there are $\lambda_1, ... , \lambda_n \in \KK $ and $y= \sum_1^n \lambda_i x_i$ such that $l = ev_{y}$.
\qed

As $ev$ is bicontinuous, we have:

\begin{prop}
\label{duality}
The function $ev$ is a linear homeomorphism from $E_w$ to $E''_w$.  \qed
\end{prop}
As a consequence of this result, our decision to put the \wstar topology on $E'$ makes it an object of $\Weak$, without any further operation on its topology: 

\begin{cor}
$E'$ is linearly homeomorphic to $(E')_w$ through the identity function $\mathrm{Id}: E' \rightarrow (E')_w$. 
\end{cor}

\proof
The topology of $E'$ is the \wstar topology induced by $E$, and the topology of  $(E')_w$ is the weak topology induced by $E'' \sim E$. 
\qed
This theory of weak and \wstar topology fits in the more general theory of dual pairs \cite[Chapter 8]{Jar81}. 

\section{Multiplicative connectives}

We will now use the evaluation function from $E$ to other domains. Indeed, consider $\mathcal{F}(E,\mathbb{C})$ some vector space of functions between $E$ and $\mathbb{C}$ and $ev: E \rightarrow \mathcal{F}(E,\KK)'$. When $\mathcal{F}(E,\KK)$ contains only linear functions, $ev$ is linear. When $E' \subset \mathcal{F}(E,\KK)$, $ev$ is injective, as $E'$ separates the points of $E$.

\begin{nota}
\label{nota:ev}
We write the following function as $ev^{\mathcal{F}(E,\KK)}$: 
\begin{equation*}
ev: 
\left\lbrace
\begin{split}
E & \rightarrow \mathcal{F}(E,\KK)' \\
x & \mapsto ( ev_x : f \mapsto f(x))
\end{split}
\right.
\end{equation*}
\end{nota}

If there is no ambiguity in the context, we will write $ev$ for $ev^{E'} : E \mapsto E''$.

\subsection{Spaces of linear maps}

\begin{defi}
Let us denote $\lin (E,F)$ the space of all continuous linear maps between $E$ and $F$, endowed with the topology of pointwise convergence on points of $E$. 
\end{defi}

A basis for the topology of simple convergence on $\lin(E,F)$ is the collection of all $$W_{x_1,...x_n, V } = \left\lbrace l \in \lin(E,F)~|~ l(x_1)\in V, ..., l(x_n) \in V \right\rbrace$$
 where $n \in \mathbb{N}$, $x_i \in E$ and $V$ is a neighbourhood of $0$ in $F$.

The \wstar topology on $E'$ is exactly the topology of simple convergence on points of $E$, thus:

\begin{fact}
For all \lct $E$, $E' \simeq \lin(E,\mathbb{K})$. 
\end{fact}

Let us write for the moment $E \otimes F$ for the algebraic tensor product between two vector space $E$ and $F$. Later on, we will endow the tensor product with a suitable topology. 

\begin{prop}[\protect{\cite[15.3.5]{Jar81},\cite[39.7]{Kot79}}]
\label{fund_prop}
$\lin (E_w,F_w)'$ is algebraically isomorphic to $E \otimes F '$. 
\end{prop}

\proof  We will sketch here the proof due to K\"othe \cite{Kot79}, as the proof by Jarchow uses the projective tensor product\footnote{The definition of the projective tensor product is recalled after the definition \ref{cotensor} of the cotensor.}.
Consider first the space $L(E,F)$ of all linear and not necessarily continuous maps between $E$ and $F$, endowed with the topology of simple convergence on points of $E$. If we choose an algebraic basis $X $ of $E$, we have $L(E,F) \simeq \prod_{x \in X} F_x$ where $F_x$ is a copy of $F$, and where the product $\prod_{x \in X} F_x$ is endowed with the product topology. Thus $L(E,F)' \simeq (\prod_{x \in X} F_x)' \simeq \bigoplus_X F_x'$ (the dual of a cartesian product is the direct sum of the duals, see Proposition \ref{dual_prod}). Linear forms in $\bigoplus_X F_x'$ are exactly finite sums of linear forms in $F'$, each one with a different domain $F_x = \{ f(x) ~~|~~ f \in L(E,F) \}$. When we consider linear forms on $\bigoplus_X F_x'$ as elements of $L(E,F)'$, we write them as finite sums $\sum_{1 \leq i \leq n}  l_i \circ ev_{x_i}$ with $x_i \in X $ and $l_i \in F'$. Thus the following linear application is well-defined and surjective :

\begin{equation*}
\left\lbrace
\begin{split}
  E \otimes F' & \rightarrow L(E,F)' \\
\sum_{1 \leq i \leq n} (x_i \otimes l_i)  & \mapsto \sum_{1 \leq i \leq n} l_i \circ ev_{x_i}
\end{split}
\right.
\end{equation*}

K\"othe shows in detail in his proof why this morphism is injective, proving that $L(E,F)'$ is algebraically isomorphic to $E \otimes F'$.

Now let us get back to $\lin (E,F)$. This space is dense in $L(E,F)$ when it is endowed with the topology of simple convergence on $E$, as for every pairwise distinct $x_1, ... x_n \in E$ and for every open set $V$ in $F$ we can find a continuous linear map $f$ such that $f(x_i) \in V$. Indeed, without loss of generality, we suppose the family $\{ x_i \} $ free. If it is not, one can reason as follows by extracting a free family from $\{ x_i \} $. Select $y \neq 0 \in V$, and for every $i \leq n$ $l_i \in E'$ such that for every $j \leq n$ $l_i(x_j) = \delta_{i,j}$. The function $ f: x \mapsto \sum_i l_i(x)y$ is linear continuous, and sends $x_i$ to $y$. 

Thus the dual of $\lin (E,F)$ is algebraically isomorphic to the dual of $L(E,F)$, that is to $E \otimes F'$. 
\qed

This proposition allows us to write every linear function  $f \in \lin (E,F)'$ as a unique finite sum $$ f = \sum_{i=1}^n l_i \circ ev_{x_i}$$ where $l_i \in F'$ and $x_i \in E$. 

Let us now describe how linear functions behave with respect to weak topologies. 

\begin{lem}
\label{linear_weak_dual}
Functions in $\lin(E,F_w)$ are exactly the linear maps from $E$ to $F$
which, when postcomposed with any map from $F'$, results in a map
belonging to $E'$. 
\end{lem}
\proof
By definition of the weak topology on $F$, a function $f: E_w \rightarrow F_w$ is continuous if and only if for every $l \in F'$  $ f \circ l: E \rightarrow \KK$ is continuous. If $f$ is linear, this means that $ f \circ l \in E'$. 
\qed

\begin{prop}
\label{linear_weak}
For all $E$, $F$  \lct, we have $\lin(E,F_w) \simeq \lin(E_w,F_w)$. 
\end{prop}

\proof
A continuous linear map from $E_w$ to $F_w$ is continuous from $E$ to $F_w$, as the weak topology is coarser than the initial topology on $E$. Consider now $ f \in \lin (E,F_w)$. For every $l \in F'$ we have $f \circ l \in E'$, thus $ f \circ l  \in (E_w)'$ by Proposition \ref{w_wstar}. By the preceding lemma, we have $f \in \lin (E_w, F_w)$. 
\qed

\subsection{Tensor and cotensor}

Various ways exist to create a \lct from the tensor product of two \lct $E$ and $F$. That is, several topologies exist on the vector space $E \otimes F$, the most prominent in the literature being the projective topology \cite[III.15]{Jar81} and the injective topology \cite[III.16]{Jar81}.  These topologies behave particularly well with respect to the completion of the tensor product, and were originally studied in Grothendieck's thesis \cite{Gro66}.

However, we would like a topology on $E \otimes F$ that would endow $\Weak$  with a structure of symmetric monoidal closed category. This is mainly why we use the inductive tensor product \cite[I.3.1]{Gro66}. So as to define this topology, we need to mention the topological product of two \lct.

\begin{defi}
Consider $E$ and $F$ two \lct. $E \times F$ is the algebraic cartesian product of the two vector spaces, endowed with the product topology, that is the coarsest topology such that the projections $p_E: E \times F \rightarrow E$ and $p_F: E \times F \rightarrow F$ are continuous. 
\end{defi}

Neighbourhoods of $0$ in $E \times F$ are generated by the sets $U \times V$, where $U$ is a $0$-neighbourhood in $E$ and $V$ is a $0$-neighbourhood in $F$.

\begin{defi}
Let us denote by $\blin (E \times F,G)$ the space of all bilinear and separately continuous functions from $E \times F$ to $G$, and by $\blin (E \times F)$ the space of all bilinear and separately continuous functions from $E \times F$ to $\KK$. We endow it with the topology of simple convergence on $E \times F$. The vector space $\blin(E \times F,G)$ is then a \lct . 
\end{defi}

Indeed, a basis of $0$ for this topology is the family of all 
\[W_{x_1,...x_n,y_1,...,y_n, V } = \left\lbrace l \in \blin (E \times F,G)~|~ l(x_1,y_1)\in V, ..., l(x_n,y_n) \in V \right\rbrace\]
with $x_1,...x_n \in E$, $y_1,...,y_n \in F$ and $V$ a $0$-neighbourhood in $G$. These sets are absolutely convex if $V$ is, and absorbent. 
Beware that is we had wanted our space of bilinear functions to be endowe with the topology of uniform convergence on bounded sets, it would not have been enough to have separately continuous function (hypocontinuity would have been necessary \cite[Ch. 42]{Tre67}). 

\begin{prop}
Consider $E$, $F$ and $G$ three \lct, and $f$ a bilinear map from $E \times F$ to $G$. Then $f \in \blin (E\times F,G_w)$ if and only if for every $l \in G'$, $l \circ f  \in \blin(E \times F)$. \qed
\end{prop}

\begin{defi}
Consider $E$ and $F$ two \lct. We endow $E \otimes F$ with the inductive topology, which is the finest topology making the canonical bilinear map $ E \times F \rightarrow E \otimes F$ separately continuous. 
\end{defi}

\begin{prop}{\protect{\cite[I.3.1.13]{Gro66}}}
\label{dual_tensor_prod}
For every \lct G, we have $\lin ( E \otimes F, G) \sim  \blin (E \times F,G) $. Especially, $ (E \otimes F) ' \sim \blin (E \times F) $. 
\end{prop}

\proof
Let us write $B(E;F,G)$ for the vector space of all bilinear maps from $E \times F$ to $G$. As $E \times F \rightarrow E \otimes F$ is separately continuous, the canonical isomorphism $ L ( E \otimes F, G) \sim B ( E \times F ,G)$ induces an injection from $ \lin ( E \otimes F , G)$ to $ \blin (E \times F , G)$. Let us show by contradiction that this injection is onto. Consider $f \in \blin (E \times F , G)$ such that its linearisation $\tilde{f} \in L(E \otimes F, G)$ is not continuous. Let us denote $E \otimes_{\tau} F$ the vector space $E \otimes F $ endowed with the coarsest topology $\tau$ making $\tilde{f}$ continuous. Then, because $f$ is separately continuous, $E \times F  \rightarrow E \otimes_{\tau} F$ is separately continuous. Thus $\tau$ is coarser than the inductive topology. This would implies that $ \tilde{f}: E  \otimes F \rightarrow  G $ would be continuous. We have a contradiction.
\qed

\begin{prop}[Associativity of $\otimes$ in $\Weak$]
Consider $E$, $F$, and $G$ three \lct.
Then $(E_w \otimes (F_w \otimes G_w)_w)_w \simeq  ((E_w \otimes F_w)_w \otimes G_w)_w$
\end{prop}

\proof
As the algebraic tensor product is associative we have  $(E_w \otimes (F_w \otimes G_w)_w)_w \sim  ((E_w \otimes F_w)_w \otimes G_w)_w $. Let us show that the two spaces bear the same topology. The dual of the first space is  $(E_w \otimes (F_w \otimes G_w)_w)' \sim \mathcal{B}(E_w \times (F_w \otimes G_w)_w)$ according to Proposition \ref{dual_tensor_prod}. One can show as above that $\mathcal{B}(E_w \times (F_w \otimes G_w)_w)$ corresponds to the space of all trilinear separately continuous functions on $E_w \times F_w \times G_w$. Likewise, the dual of the second space is $((E_w \otimes F_w)_w \otimes G_w) \sim \mathcal{B}((E_w \otimes F_w)_w \times G_w)$, which corresponds also to the space of all trilinear separately continuous functions on $E_w \times F_w \times G_w$. Then $(E_w \otimes (F_w \otimes G_w)_w)_w $ and $  ((E_w \otimes F_w)_w \otimes G_w)_w$ are algebraically isomorphic and have the same dual, thus the same weak topology. 
\qed
The associativity mapping obviously satisfies the coherence diagrams for a monoidal category \cite[VII.1]{Mac}.

\begin{prop}
Consider $E$, $F$ and $G$ three \lct. Then we have 
$$\lin ((E_w \otimes F_w)_w , G_w) \sim \blin (E_w \times F_w , G_w).$$
\end{prop}

\proof
A map $f$ lies in $\lin ((E_w \otimes F_w)_w , G_w)$ if and only if for every $l \in G'$, $l \circ f \in (E_w \otimes F_w)'$. But according to Proposition \ref{dual_tensor_prod}, we have $(E_w \otimes F_w)' \sim \blin(E_w \times F_w)$. Thus $f \in  \lin ((E_w \otimes F_w)_w , G_w)$ if and only if  the bilinear map corresponding to $f$ is in $\blin (E_w \times F_w,G_w)$.
\qed

\begin{prop}
\label{alg_mono_closed_2}
Consider $E$, $F$ and $G$ three \lct. Then we have $$\blin (E_w \times F_w,G_w) \sim \lin( E_w, \lin(F_w,G_w)_w).$$
\end{prop}

\proof
Remember from Proposition \ref{fund_prop} that $\lin(F_w,G_w)' \sim F \otimes G'$. Consider $g$ a continuous linear function from $E_w$ to $\lin(F_w,G_w)_w$. As the codomain of $g$ is $\lin(F_w,G_w)$, we have that for $x \in E$ fixed, for all $l \in G'$, $ y \mapsto l(g(x)(y))$ is continuous. So as to be continuous $g$ must satisfy that for $y$ and $l \in G'$ both fixed, $x \mapsto l(g(x)(y))$ is continuous. Consider $l \in G'$ fixed. We see that $ l \circ g$ transforms into a separately continuous map in $\blin(E_w \times F_w)$. Thus $g$ can be seen as a function $\tilde{g}$ in $\blin (E_w \times F_w,G_w)$.  The transformation of a map in $\blin (E_w,F_w \times G_w)$ into a map of $\lin( E_w, \lin(F_w,G_w)_w)$ is done likewise. 
\qed

Thus we have an algebraic isomorphism between $\lin( E_w, \lin(F_w,G_w)_w)$ and $\lin ((E_w \otimes F_w)_w , G_w)$. To show that they bear the same weak topology, we just have to show that they have the same dual. But according to Proposition \ref{fund_prop}, $\lin( E_w, \lin(F_w,G_w)_w)' \sim E_w \otimes \lin(F_w,G_w)' \sim E_w \otimes F_w \otimes G_w' \sim \lin ((E_w \otimes F_w)_w , G_w)'$. 

\begin{thm}\label{mono_closed}The category $\Weak$ is monoidal closed, as we have for each \lct $E_w$, $F_w$, $G_w$: 
$\lin( E_w, \lin(F_w,G_w)_w) \simeq \lin ((E_w \otimes F_w)_w , G_w)_w.$ \qed
\end{thm}

\begin{defi}
\label{cotensor}
The co-tensor $\parr$ of linear logic is interpreted by $E \parr F  \simeq \blin (E',F')$. 
\end{defi}

\begin{prop}
The $\parr$ connective preserves the weak topology: indeed, for every \lct $E$ and $F$, $(E \parr F)_w \simeq E_w \parr F_w$. 
\end{prop}

\proof
As $E \parr F \simeq (E' \otimes F')'$, the result follows immediately from Proposition \ref{w_wstar}. 
\qed

\subsection{A $\ast$-autonomous category}

According to Theorem \ref{mono_closed},  $\Weak$ is a monoidal closed category, with $ev_E: E \rightarrow E'' \simeq \lin (E, \lin (E , \KK)) $ being an isomorphism in this category for every object $E$. The use of weal topologies gives use a model of the classical part of linear logic, that is a $\ast$-autonomous category \cite{Bar79}.

\begin{thm}
\label{star-aut}
$\Weak$ is a $\ast$-autonomous category, with dualizing object $\KK$. 
\end{thm}

\proof
Let us take $\mathbb{K} = \bot = 1 $ the dualizing object. Then the evaluation map 

$$ (A \multimap \bot ) \otimes A  \rightarrow \bot $$

\noindent leads by symmetry of $\otimes$ and closure exactly to $ev: A \rightarrow ((A \multimap \bot ) \multimap \bot $, that is $ ev: A \rightarrow A ''$. As shown in Proposition \ref{duality}, $ev: A \rightarrow A ''$ is an isomorphism in the category $\Weak$, and $\Weak$ is $\ast$-autonomous. 
\qed

\section{Additive connectives}
\label{sec:mall}

The additive connectives of linear logic are of course interpreted by the binary product and co-product between \lct. Sadly, finite product and co-product coincide. However, they behave differently with respect to weak topology: the product preserves the weak topology, while the coproduct doesn't. See Proposition \ref{weak_prod_dual} and Section \ref{sec:polarities} for an interpretation of this phenomenon in terms of polarities. 

Practically all the results in this section are classical results from functional analysis. We nonetheless detail their proofs, which can also be found in the literature \cite{Jar81, Sch71, Kot69}. 

\begin{defi}
Consider $I$ is a set, and for all $i \in I$ $E_i$ a \lct. We define $\prod_{i \in I} E_i$ as the vector space product over $I$ of the $E_i$, endowed with the coarsest topology on $E$ such that all $p_i$ are continuous. 
\end{defi}

If $\mathcal{U}_j$ is a  basis of $0$-neighbourhoods in $E_j$, then a subbasis for the topology on $\prod_i E_i $ consists of all the 

\begin{center}
$U = U_{i_0} \times \prod_{i \in I, i \neq i_0} E_i $
 with $U_{i_0} \in \mathcal{U}_{i_0}$. 
\end{center}

\begin{defi}
We define $E \simeq \bigoplus_{j \in J} E_i$ as the algebraic direct sum of the vector spaces $E_i$, endowed with the finest locally convex topology such that every injection $I_j: E_j \rightarrow E$ is continuous. Remember that the algebraic direct sum $E$ is the subspace of $\prod_j E_j$ consisting of elements $(x_j)$ having finitely many non-zero $x_j$. 
\end{defi}

If $\mathcal{U}_j$ is a $0$-basis in $E_j$, then a $0$-basis for $ \bigoplus_j E_j$ \cite[4.3]{Jar81} is described by all the sets:
\[\textstyle U = \bigcup_{n=1}^{\infty} \sum_{k=1}^n \bigcup_j U_{j,k}
\quad\mbox{with}\quad U_{j,k} \in \mathcal{U}_j\enspace,\enspace
j \in J\enspace,\enspace k \in \mathbb{N}\,.\]
Note that this topology is finer than the topology induced by $\prod E_i$ on $\bigoplus E_i$. 

\begin{prop}[\protect{\cite[4.3.2]{Jar81}}]
\label{finite_prod}
 $I$ is finite if and only if the canonical injection from $\bigoplus_{i \in I} E_i $ to $ \prod_{i \in I} E_i$ is surjective. \qed
\end{prop}

\begin{prop}
\label{dual_prod}
For any index $I$ and all \lct $E_i$ $(\bigoplus_{i \in I} E_i)' \sim \prod_{i \in I} E'_i$ and $(\prod_{i \in I} E_i)' \sim \bigoplus_{i \in I} E'_i$. \qed
\end{prop}

\begin{prop}[\protect{\cite[II.8.8 Theorem 5 and Theorem 10]{Jar81}}]
\label{weak_prod}
We have always $(\prod_{i \in I} E_i)_w \simeq \prod_{i \in I} (E_i)_w$, but $(\bigoplus_{i \in I} E_i)_w \simeq \bigoplus_{i \in I} (E_i)_w$ holds only when $I$ is finite. 
\end{prop}

\proof
Let us show first that $(\prod_{i \in I} E_i)_w \simeq \prod_{i \in I} (E_i)_w$. The topology of $(\prod_i E_i)_w$ is the coarsest one of $\prod E_i$ making all elements of $ (\prod E_i)'$ continuous. But as $(\prod E_i)' \sim \bigoplus E_i'$, every $l \in (\prod E_i)'$ is continuous from $ \prod (E_i)_w$ to $\KK$. Thus the topology of $\prod (E_i)_w$ is finer than the topology of $(\prod_{i \in I} E_i)_w$. The topology of $\prod (E_i)_w$ is the coarsest one making all the projections $p_{i,w}: \prod (E_i)_w \mapsto (E_i)_w$ continuous. But as the  $p_i: \prod E_i  \mapsto E_i$ are continuous, all the $p_i: (\prod E_i)_w \mapsto (E_i)_w$ are continuous. Thus the topology of $(\prod_{i \in I} E_i)_w$ is finer than the topology of $\prod (E_i)_w$, and the two are equal. 

When $I$ is finite, Proposition \ref{finite_prod} and the result above tells us that $(\bigoplus_{i \in I} E_i)_w \simeq \bigoplus_{i \in I} (E_i)_w$. Suppose that $I$ is not finite, and that $E_i \neq \{ 0 \}$ for all $i$. To follow the proof by Jarchow we introduce the notion of equicontinuity: a set $B$ of linear continuous functions from $E$ to $F$ is equicontinuous if for every $0$-neighbourhood $V$ in $F$ there is a $0$-neighbourhood $U$ in $E$ such that $B(U) \subset V$. One can check as in Proposition \ref{duality}, and thanks to Lemma \ref{lem_alg} that equicontinuous subsets of $(F_w)'$ are the finite ones. Thus if $B_i$ is a finite but nonempty subset of $E_i'$, $\prod B_i$ is equicontinuous in $(\bigoplus_{i \in I} (E_i)_w)'$, but not in $((\bigoplus_{i \in I} E_i)_w)'$ as it is not finite dimensional. Thus $(\bigoplus_{i \in I} E_i)_w \nsimeq \bigoplus_{i \in I} (E_i)_w$.
\qed

We can now characterize the dual of a product and of a coproduct in the category $\Weak$.

\begin{prop}
\label{weak_prod_dual}
We have always $(\bigoplus_{i \in I} E_i)' \simeq \prod_{i \in I} E'_i$ but $(\prod_{i \in I} E_i)' \simeq \bigoplus_{i \in I} E'_i$ holds only when $I$ is finite. \qed
\end{prop}

\section{A quantitative model of linear logic}

\subsection{Quantitative semantics}

Introduced by Girard \cite{Gir88}, quantitative semantic refine the analogy between linear functions and linear programs (consuming exactly once their input). Indeed, programs consuming their resources exactly $n$-times are seen as monomials of degree $n$. General programs are seen as the disjunction of their executions consuming $n$-times their resources. Mathematically, one can apply this semantic idea by interpreting non-linear proofs as sums of n-monomials. 

The structure presented here is very simple, our spaces providing us with practically no tools except the Hahn-Banach theorem. In particular, as they satisfy no completeness condition, the notion of converging power series is not available.  Power series are converging sums of monomials, and convergence in topological vector spaces is mainly possible thanks to completeness\footnote{It appears that the weakest completeness condition necessary to model quantitative linear logic should be Mackey completeness \cite{KT15}.}. This is why we simply chose to represent non-linear maps as finite sequences over $\mathbb{N}$ of $n$-monomials. We also explore another possible exponential, inspired by what happens in the theory of formal power series, in Section \ref{subsec:formal_pws}.

We have no difficulties in defining $n$-linear mappings $\hat{f}$ on topological vector spaces. From them, we define $n$-monomials as the functions matching $x$ to $\hat{f}(x,...,x)$. Note however that to construct monomials concretely, we need a ring structure on our topological vector space, that is we need an algebra structure. We don't restrict ourselves to topological algebras, as they are particular cases of our spaces.

The exponential we define here has a lot of similarities with the free symmetric algebra studied by Mellies, Tabareau and Tasson \cite{MTT09}. The difference here is that we considered sequences of monomials in the co-Kleisli category and not $n$-linear symmetric maps. Therefore our exponential is the direct sum over $n \in \mathbb{N}$ of the dual spaces of the spaces of $n$-monomials, and not a direct sum of symmetric $n$-tensor product of $A$. 

\subsection{The exponential}

\begin{defi}
$\nlin_s (E,F)$ is the space of symmetric $n$-linear separately continuous functions from $E$ to $F$,  We write $\lin_s(E,F)$ for the space of all symmetric $n$-linear maps from $E$ to $F$.
\end{defi}

An $n$-monomial from $E$ to $F$ is a function $f: E \rightarrow F $ such that there is $\hat{f} \in \nlin (E,F)$ verifying that for all $x \in E$ $f(x)= \hat{f} (x,...,x)$. It is symmetric when for every permutation $ \sigma \in \mathbf{S_n}$, for every $x_1, .. x_n \in E$ we have $f ( x_{\sigma(1)}, ... , x_{\sigma(n)}) = f(x_1, ..., x_n)$.

\begin{prop}[\protect{The Polarization formula \cite[7.13]{KriMi}}]
\label{polar_formula}
Consider $f$ a $n$-monomial from $E$ to $F$. Then we have $f(x)= \hat{f} (x,...,x)$ where $\hat{f}$ is a symmetric $n$-linear function from $E$ to $F$ defined by: 
\begin{center}
For every $x_1,...x_n \in E$,  $\hat{f}(x_1,...,x_n)= \frac{1}{n\oc } \sum_{\epsilon_1,...,\epsilon_n=0}^1 (-1)^{n-\sum_j \epsilon_j} f ( \sum_j \epsilon_j x_j) .$
\end{center}
\end{prop}

\proof
Let us write for the multinomial coefficient : $$ {n \choose k_1, k_2, \ldots, k_m}
 = \frac{n\oc }{k_1\oc \, k_2\oc  \cdots k_m\oc } 
 = {k_1\choose k_1}{k_1+k_2\choose k_2}\cdots{k_1+k_2+\cdots+k_m\choose k_m} \,.$$  
 For every $x_1, ... , x_n $ , we have 
 $$ f ( \sum_{j=1}^n x_j) = \sum_{j_1+...+j_n = n} {n \choose k_1, k_2, \ldots, k_n} \hat{f}(\underbrace{x_1,...,x_1}_{k_1 \text{ times}}, ..., \underbrace{x_n,...,x_n}_{k_n \text{ times}}) $$
Thus 
$$ f ( \sum_j \epsilon_j x_j) = \sum_{j_1+...+j_n = n} \epsilon_1^{k_1} ... \epsilon_n^{k_n} {n \choose k_1, k_2, \ldots, k_n} \hat{f}(\underbrace{x_1,...,x_1}_{k_1 \text{ times}}, ..., \underbrace{x_n,...,x_n}_{k_n \text{ times}}) $$
and 
\begin{equation*}
\begin{split}
 & \frac{1}{n\oc } \sum_{\epsilon_1,...,\epsilon_n=0}^1 (-1)^{(n-\sum_j \epsilon_j)} f ( \sum_j \epsilon_j x_j)  \\   
& =  \sum_{\epsilon_1,...,\epsilon_n=0}^1    \sum_{j_1+...+j_n = n}  (-1)^{(n-\sum_j \epsilon_j)} \epsilon_1^{k_1} ... \epsilon_n^{k_n} \frac{1}{j_1\oc  \cdots j_n\oc } \hat{f}(\underbrace{x_1,...,x_1}_{k_1 \text{ times}}, ..., \underbrace{x_n,...,x_n}_{k_n \text{ times}}) \\  
  & = \sum_{j_1+...+j_n = n}   \frac{1}{j_1\oc  \cdots j_n\oc } \hat{f}(\underbrace{x_1,...,x_1}_{k_1 \text{ times}}, ..., \underbrace{x_n,...,x_n}_{k_n \text{ times}})    \sum_{\epsilon_1,...,\epsilon_n=0}^1  (-1)^{(n-\sum_j \epsilon_j)} \epsilon_1^{k_1} ... \epsilon_n^{k_n} \\
 \end{split}
\end{equation*}

\noindent Let us show that $\sum_{\epsilon_1,...,\epsilon_n=0}^1   (-1)^{n-\sum_j \epsilon_j} \epsilon_1^{k_1} ... \epsilon_n^{k_n}$ is non-zero if and only if $k_1=\cdots=k_n=1$. Indeed, if there is an $i$ such that $k_i \neq 1$, then there is $j$ such that $k_j = 0$, as $k_1 + ... + k_n = n$. Let us suppose $k_1=0$. Then 
\begin{equation*}
\begin{split}
\sum_{\epsilon_1,...,\epsilon_n=0}^1   (-1)^{(n-\sum_j \epsilon_j)}
\epsilon_1^{k_1} ... \epsilon_n^{k_n} 
& =  \sum_{\epsilon_2,...,\epsilon_n=0}^1   
(-1)^{(n-1- \epsilon_2 - ... - \epsilon_n)} 
\epsilon_2^{k_2 } ... \epsilon_n^{k_n}  \\ & +  
 \sum_{\epsilon_2,...,\epsilon_n=0}^1   (-1)^{n-\epsilon_2 - ... - \epsilon_n} \epsilon_2^{k_2 } ... \epsilon_n^{k_n}  \\
  & = 0 \\
 \end{split}
\end{equation*}

\noindent Thus $$\frac{1}{n\oc } \sum_{\epsilon_1,...,\epsilon_n=0}^1
(-1)^{(n-\sum_j \epsilon_j)} f ( \sum_j \epsilon_j x_j) =
\hat{f}(x_1,...,x_n)\,.
\rlap{\hbox to 100 pt{\hfill\qEd}}$$

\begin{defi}
Let us write $\nmon(E,F)$ for the space of $n$-monomials over $E$ endowed with the topology of simple convergence on points of $E$. For every \lct $E$ and $F$, $\nmon(E,F)$ is a \lct. 
\end{defi}

As a consequence of the previous proposition, we know that there is is a unique symmetric $n$-linear map $\hat{f}$ associated to a $n$-monomial $f$.
\begin{cor}
There is bijection between $\nmon(E,F)$ and $\nlin_s (E,F)$. 
\end{cor}

As we will endow $\nmon(E,F)$ with its weak topology, we need to get a better understanding of its dual. To do so, we retrieve information from the dual of $\lin^n_s(E,F)$.

\begin{prop}
\label{simeq_nmon_nlin}
For every \lct $E$ and $F$, for every $n \in \mathbb{N}$, we have $$\nmon(E,F) \simeq \lin^n_s(E,F). $$
\end{prop}

\proof
The algebraic isomorphism between the two vector spaces follows from the previous corollary, as the function mapping a $n$-linear symmetric mapping to the corresponding $n$-monomial is clearly. As they are respectively endowed with the topology of pointwise convergence of points of $E$ (resp $E \times ... \times E$), this mapping is bicontinuous. 
\qed



If we write $E_{s}^{\otimes^n}$ for the symmetrized $n^{\text{th}}$-tensor product of $E$ with himself, we have $$\nlin_s ( E, F)_w \simeq \lin (E_{s}^{\otimes^n}, F ).$$ 

\noindent As also we have $\nmon (E_w, F_w) \simeq \nmon (E,F_w) \simeq \nlin_s(E,F_w)$ by Proposition \ref{simeq_nmon_nlin}, the dual of $\nmon(E,F)'$ is the dual of $\lin (E_{s}^{\otimes^n}, F )$. Proposition \ref{fund_prop} thus gives us a way to compute it :

\begin{prop}
For every \lct $E$ and $F$, $\nmon (E, \KK)' = E_{sym}^{\otimes^n} \otimes F'$. That is, every continuous linear form $\theta$ on $\nmon (E,F)$ can be written as a finite sum of functions of the type $l \circ ev_{x_1 \otimes ... \otimes x_n}$ with $l \in F'$ and $x_1, ...x_n  \in E$.
\end{prop}

From this, we deduce that $\nmon (E_w,F_w)$ is a weak space: it is already endowed with its weak topology.

\begin{cor}
\label{nmon_weak_space}
For every \lct $E$ and $F$, we have that $\nmon (E_w,F_w)_w \simeq \nmon (E,F_w) \simeq \nmon (E_w,F_w)$.
\end{cor}

\proof
The topology on $\nmon (E_w,F_w)$ is the topology of simple convergence on $E_{sym}^{\otimes^n}$, with weak convergence on $F$. This is exactly the topology induced by its dual $E_{sym}^{\otimes^n} \otimes F'$.
\qed

\subsubsection*{The exponential}

The exponential $\oc  : \Weak  \rightarrow \Weak$ is defined as a functor on the category of linear maps. Suppose we want non-linear proofs $ E \Rightarrow F$ to be interpreted as some space of functions $\mathcal{F}(E,F)$. As the category of weak spaces and these functions is the co-Kleisli category $\Weak^{\oc }$, we have:  

\begin{equation*}
\begin{split}
(\oc E)_w & \simeq ((\oc E)_w)''\\
 & \simeq \lin (\oc  E, \KK)' \\
 & \simeq \mathcal{F} ( E, \KK)' \\
\end{split}
\end{equation*}
As we want our non-linear proofs to be interpreted by sequences of monomials, the definition of $\oc E$ is straightforward.

\begin{defi}
Let us define $\oc E$ as the \lct $\bigoplus_{n \in \mathbb{N}} \nmon (E, \KK)' $.
\end{defi}

As usual, we need to endow $\oc E$ with its weak topology.

\begin{prop}
We have $(\oc E)' \sim \prod_n \nmon (E, \KK)$, and thus $(\oc E)_w \simeq (\prod_n \nmon (E, \KK))'$.
\end{prop}

\proof
According to Proposition \ref{dual_prod}, we have that $$(\oc E)' \sim \prod_n \nmon (E, \KK) '' \sim \prod_n \nmon (E, \KK) .$$ Thus, $ (\oc E)' \simeq (\prod_n \nmon (E, \KK))_w  $, as both spaces in this equality are endowed by the topology of pointwise convergence on $\oc E$. Then $(\oc E)' \simeq \prod_n \nmon (E, \KK)_w \simeq \prod_n \nmon (E, \KK) $. Taking the dual of these spaces, we get $ \oc E_w \simeq (\prod_n \nmon (E, \KK))' $. 
\qed

As in spaces of linear functions, see Proposition \ref{linear_weak}, we have always that $\nmon (E, F_w) \simeq \nmon (E_w,F_w)$. Thus $\oc (E_w) \simeq \bigoplus_{n \in \mathbb{N}} \nmon (E_w, \KK)' \simeq \bigoplus_{n \in \mathbb{N}} \nmon (E, \KK)' \simeq \oc E$. 

\begin{nota}
We will write without any ambiguity $\oc E$ for $\oc (E_w)$ and $\oc E_w$ for $(\oc E)_w$. 
\end{nota}

\begin{defi}
For $ f \in \lin ( E_w,F_w)$ we define 
\begin{equation*}
\oc f: 
\left\lbrace
\begin{split}
\oc E_w & \rightarrow \oc F_w \\
\phi & \mapsto ( (g_n) \in \prod_{n} \nmon (F, \KK) \mapsto \phi ( (g_n \circ f)_n)
\end{split}
\right.
\end{equation*}
\end{defi}

\noindent This makes $\oc $ a functor on $\Weak$. We then endow $\oc $ with its co-monadic structure. The structure of the co-monad is based on the correspondence between $ f \in \prod_m \mmon (E,F)$ and $g \in \prod_n \nmon (E,G)$, that is $ f \circ g  \in \prod_p \pmon (E,G)$ with $$ (f \circ g)_p = \sum_{k |p} g_k \circ f_{\frac{p}{k}} . $$


\begin{rem}
At this point we must pay attention to the arithmetic employed here. So as to avoid infinite sums and a diverging term for $(f \circ g)_0$\footnote{The problem of the possible divergence of the nonzero term can be found also in the theory of formal power series \cite[IV.4]{Hen74}, where composition is only allowed for series with no constant component.}, we allow for only $0$ to divide $0$. Thus $(f \circ g)_0 = f_0 \circ g_0$.   
\end{rem}

\begin{prop}
\label{comonad}
The functor $ \oc : Lin \rightarrow Lin$ is a co-monad. Its co-unit $\epsilon: \oc  \rightarrow 1$ is defined by 
\begin{equation*}
\epsilon_E 
\left\lbrace
\begin{split}
\oc E_w & \rightarrow E_w \\
\phi  & \mapsto \phi_1 \in E'' \simeq E
\end{split}
\right.
\end{equation*}
The co-unit is the operator extracting from $\phi \in \oc E$ its part operating on linear maps. The co-multiplication $\delta: \oc  \rightarrow \oc  \oc  $ is defined as : 

\begin{equation*}
\delta_E 
\left\lbrace
\begin{split}
\oc E_w \simeq (\prod_p \pmon ( E, \KK ))' & \rightarrow \oc \oc E_w \simeq \left( \prod_n \nmon ( [ \prod_m \mmon(E, \KK)]', \KK) \right)' \\
\phi \in (\prod_p \pmon ( E, \KK ))'  & \mapsto \left[ (g_n)_n  \mapsto \phi ( \left( x \in E \mapsto \sum_{k|p} g_k [ (f_m)_m  \mapsto f_{p|k} (x) ] \right)_p) \right]
\end{split}
\right.
\end{equation*}
As explained before, we want to have on our co-Kleisli category a composition such that $ (f \circ g)_p = \sum_{k |p} g_k \circ f_{\frac{p}{k}} . $ The co-multiplication $\delta: \oc  \rightarrow \oc  \oc  $ can be seen as a continuation-passing style transformation of this operation. Indeed, consider $\phi \in \oc E$. We construct $\delta ( \phi)$ as a function in $ (\prod_n \nmon (\oc E, \KK))'$   mapping a sequence $(g_n)_n$ to $\phi$ applied to the sequences of $p$-monomials on $E$ defined as $$x \in E \mapsto \sum_{k|p} g_k [ (f_m)_m  \mapsto f_{p|k} (x) ].$$
\end{prop}\smallskip

\label{continuation}
\noindent So as to show that $\oc $ is in fact a co-monad, we need to understand better the elements of $ \oc E $.  The space $\oc E $ is defined as $\oplus_n \nmon(E, \KK)'$, so $\phi \in \oc E$ can be described as a finite sum $ \phi = \sum_{n=1}^N  \phi_n$ with $\phi_n \in \nmon (E, \KK)'$. 
The proofs presented below are based more on the idea of non-linear continuations than on a combinatoric point of view. The space $\oc E_w =  (\prod_p \pmon ( E, \KK ))'$ can be thought of as a space of quantitative-linear continuations, $\KK$ being the space of the result of a computation. Indeed, if $x$ is a program of type $E$, its continuation $k$ is of type $\prod_p \pmon ( E, \KK )$, and the type of the continuation passing-style transformation $ \lambda k. k x$ of $x$ is $(\prod_p \pmon ( E, \KK ))'$, as $ \lambda k. k x$ is linear in $k$.

\proof
We have to check the two equations of a co-monad, that is:
\begin{itemize}
\item $ \delta \oc  \delta = \oc  \delta \delta $
\item $ \epsilon_{\oc } \delta = \oc  \epsilon \delta = Id_{\oc }$
\end{itemize}
Let us detail the computation of the equation. Remember that we write $ev^{\nmon(F, \KK)}$ for $ev : F  \mapsto \nmon(F, \KK)'$.  For every $ \phi = \sum \phi_p  \in \oc E$, we have:

\begin{equation*}
\begin{split}
\epsilon_{\oc E} \delta_E ( \phi) & 
  = \epsilon_{\oc E} \left(  (g_n)_n \in \prod \nmon( \oc E, \KK) \mapsto \phi  \left( [x \in E \mapsto \sum_{k|p} g_k ( (f_m)_m \in \oc E \mapsto f_{\frac{p}{k}} (x) ) ]_p \right)    \right)  \\
  & = \epsilon_{\oc E} \left(  (g_n)_n  \mapsto \phi \left( [  x \in E \mapsto \sum_{k|p} g_k ( ev_x^{\mathcal{H}^{p/k}(E, \KK)}) ]_p \right)   \right) \\
  & =  \epsilon_{\oc E}  \left(  (g_n)_n  \mapsto \sum_p \phi_p ( x \mapsto \sum_{k|p} g_k ( ev_x^{\mathcal{H}^{p/k}(E, \KK)}  ))  \right) 
  \end{split}
\end{equation*}
 As $\epsilon_{\oc E}$ maps a function in $\oc E_w \simeq (\prod_n \nmon (E, \KK))'$ to its restriction to  $\lin(E, \KK)$, and then to the corresponding element in $\oc E$, we have without using the isomorphism $\oc E'' \simeq \oc E $:
 
\begin{equation*}
\begin{split}
\epsilon_{\oc E} \delta_E ( \phi) &  =   g_1 \in \oc E' \mapsto \sum_p \phi_p ( x \mapsto \sum_{k|p} \underbrace{g_k ( ev_x^{\mathcal{H}^{p/k}(E, \KK)} )) }_{\neq 0 \text{ if and only if } k = 1 }   \\
  & =  g_1 \in \oc E' \mapsto \sum_p \phi_p ( x \mapsto g_1 ( ev_x^{\mathcal{H}^{p}(E, \KK)} )) 
\end{split}
\end{equation*}
As $g_1$ lives in $\oc E'  \simeq \prod_m  \mmon(E, \KK)$, we can write $g_1$ as a sequence $(g_{1,m})_m$ of  $m$-monomials: 

\begin{equation*}
\begin{split}
\epsilon_{\oc E} \delta_E ( \phi)&  =   g_1 \in \oc E' \mapsto \sum_p \phi_p (x \mapsto ev_x^{\mathcal{H}^{p/k}(E, \KK)} ( g_{1,p}))    \\
  & =   g_1 \in \oc E'  \mapsto \sum_p \phi_p (x \mapsto  g_{1,p}(x))  \\
  & = g_1 \in \oc E'  \mapsto \sum_p \phi_p ( g_{1,p}) \\
   & = g_1 \in \oc E'  \mapsto \phi (g_1)
\end{split}
\end{equation*}
With the isomorphism $\oc E'' \simeq \oc E $ we obtain  $\epsilon_{\oc } \delta = Id_{\oc }$. 

\noindent The equation $ \oc  \epsilon \delta = Id $ is proved likewise: consider 
$\phi = \sum \phi_p  \in \oc E$. Then

\begin{equation*}
\begin{split}
\oc  \epsilon \delta (\phi) & = \oc  \epsilon \left(  (g_n)_n  \in \prod \nmon( \oc E, \KK) \mapsto \sum_p \phi_p ( x \mapsto \sum_{k|p} g_k ( ev_x^{\mathcal{H}^{p/k}(E, \KK)} )  )\right)  \\
& =  (h_m)_m \in \prod \mmon ( E,\KK) \mapsto \sum_p \phi_p ( x \mapsto \sum_{k|p} h_k \circ \underbrace{\epsilon ( ev_x^{\mathcal{H}^{p/k}(E, \KK)} )}_{ \neq 0 \text{ if an only if } \frac{p}{k}=1})  \\
& = (h_m)_m \in \prod \mmon ( E,\KK) \mapsto \sum_p \phi_p ( x \mapsto h_p (x))  \\
& =  (h_m)_m \mapsto \phi ( (h_m)_m ) \\
& = \phi 
\end{split}
\end{equation*}
So $ \oc  \epsilon \delta = Id $.
%
%
%
\qed

\begin{prop}
\label{exp_monoidal}
This co-monad is also symmetric monoidal. We have a natural transformation :
\begin{align*}
\mu_{E,F} : &  \oc E \otimes \oc F \rightarrow \oc (E \otimes F) \\
            &  \phi \otimes \psi \mapsto \left( (f_n)_n \mapsto \phi \left( ( x \mapsto ( \psi (y \mapsto f_n (x \otimes y) ) )_n \right)  \right) 
\end{align*}
 and a morphism 
\begin{align*}
\mu   & : \CC \rightarrow \oc \CC \\
       & t \mapsto ev_t
\end{align*}
which statisfies the required commutation diagrams.
\end{prop}

\begin{defi}
\label{whynot}
The $\wn$ connective of linear logic is interpreted as the dual of $!$, that is $$ \wn E \simeq  ( \oc E)'  \simeq  \prod_n \nmon(E', \KK).$$
\end{defi}\medskip

\noindent We will write $\Weak^{\oc }$ for the co-Kleisli category of $\Weak$ with $\oc $. We first show that morphisms of this category are easy to understand, as they are just sequences of $n$-monomials.

\subsection{The co-Kleisli category}

The exponential above was chosen because of its co-Kleisli category. Indeed, we want to decompose non-linear proofs as disjunctions of $n$-linear proofs, and the simplest way to do that is to interpret non-linear maps from $E$ to $F$, that is linear maps from $\oc E$ to $F$, as sequences of $n$-monomials from $E$ to $F$.

\begin{thm}
\label{adj}
For all \lct $E$ and $F$, $\lin (\oc E_w, F_w) \sim \prod_{n \in \mathbb{N}} \nmon (E_w, F_w) $. 
\end{thm}

\proof
Consider $f \in \lin (\oc E_w, F_w)$. 
Define, for each $n \in \mathbb{N}$,  $ f_n: x \in E_w \mapsto f(ev_x^{\nmon (E_w, \KK)})$. 
Then $f_n$ is clearly $n$-linear. 
Let us show that it is continuous from $E_w$ to $F_w$. 
Consider $l \in F'$. 
Then $x \mapsto ev_x^{\nmon (E_w, \KK)}$ is continuous from $E_w$ to $\nmon (E_w, \KK)'$, as the latter space is endowed with the topology of simple convergence on $\nmon (E_w, \KK)$.  The injection $\nmon (E_w, \KK)' \hookrightarrow \oc E_w $ is continuous, as $ \oc E_w \simeq ( \bigoplus_k  \mathcal{H}^k(E_w, \KK)' )_w$ according to Proposition \ref{weak_prod_dual}, 
and as $ \nmon (E_w, \KK)' \hookrightarrow \bigoplus_k  \mathcal{H}^k(E_w, \KK)'$ and $\bigoplus_k  \mathcal{H}^k(E_w, \KK)' \hookrightarrow ( \bigoplus_k  \mathcal{H}^k(E_w, \KK)' )_w$ are continuous. The following $n$-monomial is continuous:  

$$ f_n: E_w \overset{ev}{\longrightarrow} \nmon (E_w, \KK)' \hookrightarrow \bigoplus_k  \mathcal{H}^k(E_w, \KK)' \hookrightarrow ( \bigoplus_k  \mathcal{H}^k(E_w, \KK)' )_w \overset{f}{\longrightarrow} F_w $$

Thus $f_n \in \nmon (E_w,F_w)$. To every $f \in \lin (\oc E_w, F_w)$
we associate in this way $(f_n) \in \prod_n \nmon (E_w,F_w) $. Consider now $(f_n) \in \prod_n \nmon (E_w,F_w)$ and define $f: \phi \in \oc E_w \mapsto ( l \in F' \mapsto \phi ( (l \circ f_n)_n))$. The function $f$ is well-defined as $l \circ f_n \in \nmon (E_w, \KK)$ for every $ n \in \mathbb{N}$ and every $l \in F'$. When $\phi$ is fixed, let us denote $\phi_f$ the function  $l \in F' \mapsto \phi ( (l \circ f_n)_n))$. Then:
\begin{itemize}
\item $l \in E' \mapsto l \circ f_n \in \nmon (E, \KK)$ is continuous as $F'$ (resp. $\nmon (E, \KK)$) is endowed with the topology of simple convergence on points of $F$ (resp. on points of $E$);
\item $ l \mapsto (l \circ f_n)_n \in \prod_n  \nmon (E_w, \KK)$ is then continuous by definition of the product topology;
\item $\phi_f$ is then continuous.
\end{itemize}

Thus $\phi_f \in F'' \simeq F$. For each $\phi$, there is $y \in F$ such that $\phi_f = ev_y$. We can now consider $f: \phi \in \oc  E \mapsto y \in F$. $f$ is clearly linear in $\phi$. It is continuous as $ \oc E_w$ is endowed with the topology of simple convergence on $\prod_{n \in \mathbb{N}} \nmon (E_w, \KK)$. 

Finally, one can check that the mappings $\theta: f \in \lin (\oc E_w, F_w)  \mapsto (f_n) \in \prod_n \nmon (E_w, F_w) $ and $\Delta: (f_n) \in \prod_n \nmon (E_w,F_w) \mapsto f  \in \lin (\oc E_w, F_w)$ just described are inverse one of the other. 
\qed

Let us show now that the isomorphism described above is a homeomorphism.

\begin{thm}
For all \lct $E$ and $F$, $$\lin (\oc E_w, F_w) \simeq \prod_{n \in \mathbb{N}} \nmon (E_w, F_w), $$ and therefore 

$$\lin (\oc E_w, F_w)_w \simeq (\prod_{n \in \mathbb{N}} \nmon (E_w, F_w))_w \simeq \prod_{n \in \mathbb{N}} \nmon (E_w, F_w)_w.$$ 
\end{thm}

\proof
Let us show first that the function $ \theta: f \in \lin (\oc E_w, F_w) \mapsto  (f_n) \in \prod_n \nmon (E_w, F_w) $ is continuous. It is enough to show that $ f \mapsto f_n$ is continuous. Consider $(f_{\gamma})_{\gamma \in \Gamma}$ a net converging towards $f$ in $\lin (\oc E_w, F_w)$. Thus for every $\phi \in \oc E_w$ $f_{\gamma}(\phi)$ converges towards $f(\phi)$ in $F_w$. For every $ x \in E $ the net $f_{\gamma}(ev_x \in \nmon (E_w, \KK)')$ converges towards $f(ev_x)$ in $F_w$ thus the net $(f_{\gamma,n})$ converges towards $f_n$ and $\theta$ in continuous.
The proof that $\Delta: (f_n) \in \prod_n \nmon (E_w,F_w) \mapsto f  \in \lin (\oc E_w, F_w)$ is continuous is done likewise.
\qed

The composition in $\Weak^{\oc }$ is thus given by the definition of a co-Kleisli category. If $f \in \lin ( \oc  E, F)$ and $g \in \lin (\oc F,G)$ we define:

$$ g \circ f: \oc E \overset{\delta_E}{\longrightarrow} \oc \oc E \overset{\oc f}{\longrightarrow} \oc F \overset{g}{\longrightarrow} G .$$

\begin{nota}
For $f \in \lin ( \oc  E, F)$, we will write $( \widetilde{f}_m)_m$ the corresponding sequences of monomials in $\prod_m \mmon (E,F)$.
\end{nota}

\begin{prop}
\label{composition}
For every $f \in \lin ( \oc  E, F)$ and $g \in \lin (\oc F,G)$, we have 
$$ \widetilde{(g \circ f)}_p = \sum_{k | p} \widetilde{g}_k \circ \widetilde{f}_{\frac{p}{k}}. $$
\end{prop}

\proof
By definition, for $\phi \in \oc E$, 
\begin{equation*}
\begin{split}
g \circ f (\phi) & = g  (\oc  f (\delta (\phi)) \\
 & = g ( (g_n) \in \prod_{n} \nmon (F, \KK) \mapsto \delta (\phi) ( (g_n \circ f)_n)
\end{split}
\end{equation*}
For every $p  \in \mathbb{N}^{\ast} $, and  $x \in E$, we have:

\begin{equation*}
\begin{split}
(\widetilde{g \circ f})_p ( x) & = g \circ f (ev_x^{\pmon(E, \KK)} ) \\
 & = g ( (g_n) \in \prod_{n} \nmon (F, \KK) \mapsto \delta (ev_x^{\pmon(E, \KK)}) ( (g_n \circ f)_n))
\end{split}
\end{equation*}
Now $\delta (ev_x^{\pmon(E, \KK)}) = (h_j)_j \in \jmon( \oc E, \KK) \mapsto \sum_{k|p} h_k ( ev_x^{\mathcal{H}^{p/k}(E, \KK)})$. Thus

\begin{equation*}
\begin{split}
(\widetilde{g \circ f})_p ( x) & = g ( (g_n) \in \prod_{n} \nmon (F, \KK) \mapsto \delta (ev_x^{\pmon(E, \KK)}) ( (g_n \circ f)_n) \\
&   =  \sum_{k|p} \tilde{g}_k ( f( ev_x^{\mathcal{H}^{p/k}(E, \KK)}  )) \\
& = \sum_{k|p} \tilde{g}_k \circ \tilde{f}_{p/k} \rlap{\hbox to 247 pt{\hfill\qEd}}
\end{split}
\end{equation*}

\subsection{Cartesian closedness}
Let us show that $\Weak^{\oc }$ endowed with the cartesian product described in Section \ref{sec:mall} is cartesian closed.

\begin{thm}
For every \lct $E$, $F$ and $G$, we have: 
$$\prod_{p \in \mathbb{N}} \pmon( E_w \times F_w, G_w) \simeq \prod_{n \in \NN} \nmon ( E_w , \prod_{m \in \NN} \mmon (F_w,G_w)) .$$
\end{thm}\medskip

\noindent The equality above means also that $$(\prod_{p} \pmon( E_w \times F_w, G_w))_w \simeq [\prod_n \nmon ( E_w , [\prod_m \mmon (F_w,G_w)]_w)]_w ,$$ as $(\prod_m \mmon (F_w,G_w))_w) \simeq \prod_m \mmon (F_w,G_w)_w$ by Proposition \ref{weak_prod}, and since $\pmon (E_w, F_w)$ is already endowed with its weak topology by Proposition \ref{nmon_weak_space}.

\proof
For every $n \in \mathbb{N}$, for every \lct $E$ and $F$ we have $\nmon (E,F)  \simeq \nlin_s (E,F)$. We are therefore going to prove that for every \lct $E$, $F$ and $G$: 
$$\prod_{p} \plin_s( E_w \times F_w, G_w) \simeq \prod_n \nlin_s ( E_w , \prod_m \mlin_s (F_w,G_w)_w) .$$
In the following, we will write $\vec{x}$ for some tuple $(x_1,...x_n)$ in $E \times ... \times E$. 
Let us fix $E$, $F$ and $G$, and define: 

\begin{equation*}
\phi:
\left\lbrace
\begin{split}
   \prod_{p} \plin_s( E_w \times F_w, G_w)  \rightarrow  & \prod_n  \nlin_s ( E_w , \prod_m  \mlin_s (F_w,G_w)_w) )\\
  (f_p)   \mapsto & [ \vec{x} \mapsto  ( \vec{y}   \mapsto f_{n+m} ((x_1,0),...,(x_n,0),(0,y_1),...,(0,y_m)) )_m ]_n 
\end{split}
\right.
\end{equation*} 

\noindent Let us show that $\phi$ is  well-defined.

\begin{itemize}
\item Consider $(f_p)_p \in \prod_{p} \plin_s( E_w \times F_w, G_w)$, $n \in \mathbb{N}$, $x \in E$, and $m \in \mathbb{N}$. Then  $$y \in F \mapsto  f_{n+m} ((x_1,0),...,(x_n,0), (0,y_1),...,(0,y_m))$$ is $m$-linear and symmetric, and continuous from $F_w$ to $G_w$ as $f_{n+m}: E_w \times F_w  \rightarrow G_w$ is continuous.

\item Consider $(f_p)_p \in \prod_{p} \plin_s( E_w \times F_w, G_w)$ and $n \in \mathbb{N}$. Then $$x_1,...,x_n \in E_w \mapsto  \left(  y_1,...,y_m \in F \mapsto f_{n+m} ((x_1,0),...,(x_n,0),(0,y_1),...,(0,y_m)) \right)$$ is clearly $n$-linear and symmetric. It is continuous from $E_w$ to $\mlin_s(F_w,G_w) $ as the latter bears the topology of simple convergence, and as $f_{n+m}$ is continuous from $E_w \times f_w$ to $G_w$. Since the weak topology on $ \mlin_s(F_w,G_w) $ is coarser than the strong topology, the function considered is also continuous from $E_w$ to $\mlin_s(F_w,G_w)_w$.
\end{itemize}
We want to define the inverse function $\psi$ of $\phi$. Thus, in
particular, $\phi$ is a function from $\prod_n \nlin_s ( E_w , \prod_m  \mlin_s (F_w,G_w))$ to $\prod_{p} \plin_s ( E_w \times F_w, G_w)$. Consider $$f_n \in \nlin_s ( E_w , \prod_m  \mlin_s (F_w,G_w)) $$ and let us write $f_{n,\vec{x},m}$ for $(f_n (\vec{x}))_m \in \mlin (F_w,G_w)$. If $p \geq \max{n,m}$, then the following function is $n+m$-linear:
\[((x_1,y_1), ..., (x_p,y_p)) \mapsto f_{n,(x_1,...,x_n),m}(y_1,...,y_m).\]
When $p$ is fixed, $\psi(f)_p$ collects all possible ways to produce a $p$-linear function as above, with $p=n+m$. As all possible permutations are considered, $\psi(f)_p$ is symmetric.

\begin{flushleft}
  \begin{equation*}
\psi:
\left\lbrace
\begin{split}
   \prod_n \nlin_s ( E_w , \prod_m  \mlin_s (F_w,G_w))    & \rightarrow  \prod_{p} \plin_s ( E_w \times F_w, G_w) \\
  [f_n:\vec{x} \mapsto (f_{n,\vec{x},m})_m]_n & \mapsto  [\overrightarrow{(x,y)}  \mapsto \sum_{\substack{ I, J \subset [| 1,p|] \\ card(I)=n \\ card(J)=m \\ n+m = p}} \frac{1}{\binom{p}{n}} f_{n,\{x_i\}_{i \in I},m} (\{y_j\}_{j\in J}) ]_p
\end{split}
\right.
\end{equation*}
 \end{flushleft} 
where in the index of the sum $I$ and $J$ are ordered subsets (that is
sequences) of $[|1,nl]$. 

  If $I=\{i_1,\dots,i_n\}$ and $J= \{ j_1, ..., j_m \}$, we use
  $f_{n,\{x_i\}_{i \in I},m} (\{y_j\}_{j\in J})$ as a shorthand notation for $f_{n,(x_{i_1},\dots,x_{i_n}),m} (y_{j_1},\dots,y_{j_m})$.

Let us show that $\psi$ is well defined. Consider

 $$[f_n: x_1,..x_n \mapsto (f_{n,\{x_i\},m})_m]_n \in  \prod_n \nlin_s ( E_w , \prod_m  \mlin_s (F_w,G_w)_w)  .$$
 
\noindent The function mapping $ ((x_1,y_1),...,(x_p,y_p)) \in (E_w \times F_w)^p$ to $ f_{n,\{x_i\}_{i \in I},m} (\{y_j\}_{j\in J})$ is $n+m$-linear and symmetric. It is continuous, as the restrictions to fixed terms in $E_w$ or $F_w$ are continuous. So $\psi$ is well defined. 
Note that both  $\phi$ and $\psi$ are continuous as the spaces $\nlin_s (E,F)_w$ are endowed with the topology induced by their dual $ E^{\otimes^n}_{sym} \otimes F' $.
Finally, one checks that $\phi$ and $\psi$ are each other's inverse. Consider $f \in \prod_p \plin_s(E_w \times F_w , G_w)$. Then $\psi ( \phi (f))$ corresponds to the function mapping $p$ to the function in $\plin_s(E_w \times F_w , G_w)$ mapping $((x_1,y_1),\dots,(x_p,y_p))$ to:

$$ \sum_{\substack{ I, J \subset [| 1,p|] \\ card(I)=n \\ card(J)=m \\ n+m = p}} \frac{1}{\binom{p}{n}} f( (x_{i_1},0), \dots,   (x_{i_n},0), (0, y_{j_1}), \dots, (0, y_{j_m})).$$

\noindent By $n$-linearity of $f_p$ this sum equals 

$$f_p((x_1,y_1), \dots, (x_p, y_p)).$$  

\noindent Thus $\psi \circ \phi = \text{Id}$. 
\noindent Consider now 

$$g = [g_n: x_1,..x_n \mapsto (g_{n,\{x_1,\dots,x_n\},m})_m]_n\in \prod_n \nlin_s ( E_w , \prod_m  \mlin_s (F_w,G_w)) .$$ 

\noindent Let us show that $ \phi (\psi (g))= g$. The function $\psi (g)$ maps $p$,  $((z_1,w_1),\dots,(z_p,w_p))$ to 
$$ \sum_{\substack{ I, J \subset [| 1,p |] \\ card(I)=a \\ card(J)=b\\ a +b =p }}  \frac{1}{\binom{p}{a}} g_{n,\{z_i\}_{i \in I},m}(\{w_j \}_{j \in J}).$$
The function $\phi( \psi (g))$ maps $n\in \NN $, $x_1, ..., x_n \in E $, $m \in \NN$ and $y_1, ..., y_m \in F$ to this function applied to $n+m$ and $((x_1,0),...,(x_n,0),(0,y_1),...,(0,y_m))$. But notice that $g_{n,\{z_i\}_{i \in I},m}(\{w_j \}_{j \in J})$ is null as soon as one of the $z_i$ or one of the $w_i$ is null. So $\phi( \psi (g))$ applied to $n\in \NN $, $x_1, ..., x_n \in E $, $m \in \NN$ and $y_1, ..., y_m \in F$  results in 
$$ \frac{1}{\binom{n+m}{n}} \sum_{\substack{ I, J \subset [| 1,n +m|] \\ I ={1,n} \\ J=\{ 1,m\} }}   g_{n,\{z_i\}_{i \in I},m}(\{w_j \}_{j \in J})$$
which is exactly  $g_{n,\{x_1,\dots,x_n\},m}(y_1,\dots,y_m)$. 
\qed

\begin{thm}
For all \lct $E$ and $F$ we have: 
$$ \oc   (E_w \times F_w)  \simeq \oc E_w \otimes \oc F_w $$
\end{thm}

\proof
This follows from the cartesian closedness of $\Weak^{\oc }$, the monoidal closedness of $\Weak$, and the description of $Weak^{\oc }$ obtained in Theorem \ref{adj}. Indeed

\begin{equation*}
\begin{split}
\oc  ( E_w \times F_w)& \simeq \prod_p \pmon (E_w \times F_w, \KK)' \\
        & \simeq \prod_n \nmon ( E_w , \prod_m  \mmon (F_w,\KK)_w))' \\
        & \simeq \lin ( \oc E_w,  \prod_m  \mmon (F_w,\KK)_w))'   \\
        & \simeq \lin (\oc E_w, \lin ( \oc F_w, \KK)_w )\\
        & \simeq ( \oc E_w \otimes \oc F_w)''\\
        & \simeq \oc E_w \otimes \oc F_w.\rlap{\hbox to 206 pt{\hfill\qEd}}
\end{split}
\end{equation*}

\subsection{Derivation and integration}

As a quantitative model of linear logic, this model interprets differential linear logic \cite{Ehr11}. However, the interpretation of derivation remains combinatorial, and not as close to the usual differentiation operation as one would wish for. 

\begin{defi}
The co-dereliction rule of differential linear logic is interpreted by:
\begin{equation*}
\text{coder}_E:
\left\lbrace
\begin{split}
E_w & \rightarrow \oc E_w \\
x & \mapsto ( (f_n) \in \prod_{n} \nmon (E, \KK) \mapsto f_1(x))
\end{split}
\right.
\end{equation*}
\end{defi}

\begin{prop}
For every space $E$, $ \text{coder}_E$ is a linear continuous function from $E_w$ to $\oc E_w$.
\end{prop}

\proof
Let us fix $\phi = ( \phi_n)_n \in (\oc E_w)' \simeq \prod_n \nmon(E, \KK)$. Then $\phi \circ coder_E$ maps $x \in E$ to $ \phi_1 (x)$. As $\phi_1 \in E'$, $\text{coder}_E$ is continuous from $E_w$ to $\oc E_w$. 
\qed

For every \lct $E$, $!E$ bears a structure of bialgebra. This reflects the symmetric structures of the exponential rules in differential linear logic. We detail this structure. 
\begin{itemize}
\item $ \Delta: \oc E \rightarrow \oc E \otimes \oc E $ interprets the contraction rules of linear logic. It is corresponds to $ \oc E \rightarrow \oc (E \times E) \simeq \oc E \otimes \oc E $, where the first computation is the functor $ \oc$ composed with the diagonalisation morphism, and the second corresponds to the Seely isomorphism. In fact, for $\phi \in !E$ and if we write $ ev_{x,i} : (f_n)_n \in \prod_n \nmon (E, \KK) \mapsto f_i (x)$ we have : 
$$ \Delta : \phi \in !E \mapsto \left( g \in ( ! E \otimes !E )' \mapsto \phi ( [ x \mapsto \sum_{i+j=n} g ( ev_{x,i} \otimes ev_{x,j}) ]_n) \right) . $$
\item $e: \oc E \rightarrow \mathbb{C} $ is $e(\phi)= \phi (1) \in S(E, \mathbb{C}) $. It interprets the  weakening rule of linear logic.
\item $ \bigtriangledown: \oc E \otimes \oc E \rightarrow \oc E$ is defined by $\bigtriangledown ( h \mapsto \phi ( x \mapsto \psi (y \mapsto h (x+y))))$. It interprets the co-contraction rule of differential linear logic.
\item The  co-weakening rule of differential linear logic is interpreted by $\nu: \mathbb{C} \rightarrow \oc E $ is $\nu (1) = ev_0$. 
\end{itemize}

Recall that $!$ is also a symmetric lax monoidal functor (see Propositions \ref{comonad} and \ref{exp_monoidal}). All these morphisms are necessary to build a differential structure.
\begin{prop}
$\Weak$ endowed with $coder$ is a differential category \cite{BCS06}.
\end{prop}
\proof
As shown by Fiore~\cite{Fiore}, it is enough to prove that the following diagrams hold :
  
\begin{itemize}

\item Strenght :

\begin{tikzpicture}
\node (a) at (-4,0) {$E \otimes \oc F$};
\node (b) at (0,0) {$\oc E \otimes \oc F$};
\node (c) at (4,0) {$\oc (E \otimes F)$};
\node (d) at (0,-2) {$E \otimes F$};

\draw[->] (a) -- (b) node [above,midway] {$coder_E \otimes Id$};
\draw[->] (b) -- (c) node [above,midway] {$\mu_{E,F}$};
\draw[->] (a) -- (d) node [left,midway] {$Id \otimes \epsilon_F$};
\draw[->] (d) -- (c) node [right,midway] {$coder_{E \otimes F}$};
\end{tikzpicture}

\item Comonad :

\begin{tikzpicture}
\node (a) at (-0,0) {$E$};
\node (b) at (2,0) {$\oc E $};
\node (c) at (0,-2) {$E$};

\draw[->] (a) -- (b) node [above,midway] {$coder_E$};
\draw[->] (b) -- (c) node [right,midway] {$\epsilon_E$};
\draw[->] (a) -- (c) node [left,midway] {$Id$};

\node (a) at (4,0) {$E$};
\node (b) at (8,0) {$\oc E $};
\node (c) at (12,0) {$\oc \oc E$};
\node (a') at (4,-2) {$E \otimes 1$};
\node (b') at (8,-2) {$\oc E  \otimes \oc E$};
\node (c') at (12,-2) {$\oc \oc E \otimes \oc \oc E$};

\draw[->] (a) -- (b) node [above,midway] {$coder_E $};
\draw[->] (b) -- (c) node [above,midway] {$\delta_E$};
\draw[->] (a) -- (a') ;
\draw[->] (c') -- (c) node [right,midway] {$\bigtriangledown$};
\draw[->] (a') -- (b') node [below,midway] {$coder_E \otimes \nu $};
\draw[->] (b') -- (c') node [below,midway] {$coder_{\oc E} \otimes \delta_E$};
\end{tikzpicture}

\end{itemize}

\noindent In our category, both branches of the strenght diagrams computes the following function : $$( x \otimes \phi) \mapsto \left( (f_n)_n \mapsto \phi ( y \mapsto f_1 ( x \otimes y) ) \right).$$ The first comonad diagram is immediate by the definition of $\epsilon$. The second diagram computes the function $$ x \in E \mapsto (g_p)_p \in \prod_p \mathcal{H}^p ( \oc E, \KK) \mapsto g_1 ( (f_n)_n \mapsto f_1 (x) )\,.\eqno{\qEd}$$\medskip

\noindent We do not have an interpretation of a syntactic integration in this category. Indeed, the existence of Ehrhard's anti-derivative operator \cite[2.3]{Ehr11} would imply some sort of integration. We do not have a way to integrate in our spaces, as no completeness condition is verified. It is noticeable that if our spaces were reflexive, that is isomorphic to their bidual when the dual is endowed with the topology of uniform convergence over bounded sets, a weak integration would be available. 

\subsection{An exponential with non-unit sequences}
\label{subsec:formal_pws}

Inspired by the substitution problem in the theory of formal power series \cite[Chapter 1]{Hen74}, we could have used another composition between sequences of monomials. Indeed, such sequences can be considered as generalized formal power series. That is, to a sequence $(f_n)_n$ corresponds a formal sum $\sum f_n$, where no notion of convergence is employed. A formal power series is a denumerable sum $A(X) = \sum_m a_m X^n$ where the $a_i$ are coefficients in some commutative ring $R$. If $B =  \sum_p B_n X^p$ is another formal power series, one has  :

\begin{equation*}
\begin{split}
B (A(x)) & = \sum_p b_p (A(x))^p
\end{split}
\end{equation*}
For every $n \in \mathbb{N}$, $A(x)^p$ can be computed as if the sum in $A$ were convergent. That is, $A(x)^p= \sum_k c_k^p X^k$ with 
$$ c_k^p = \sum_{n \geq 0} \sum_{k_1 + ... + k_n = p} a_{k_1} \times a_{k_2} \times ... \times a_{k_n}.$$
This sum is infinite, which causes a problem since no notion of convergence is employed here. However, if $A$ is non-unit, that is if $a_0=0$ this sum becomes finite : 
$$c_k^p = \sum_{p \geq n \geq 1} \sum_{m \geq k_i \geq 1 \atop k_1 + ... + k_n = p} a_{k_1} \times a_{k_2} \times ... \times a_{k_n}. $$
With the same ideas, one can construct a comonad with the a non-unit version of functor $\oc  : \Weak \rightarrow \Weak$ such that, in the co-Kleisli category $\Weak_\oc $ corresponds with substitution if its morphisms are seen as formal power series $f = \sum_{n \geq 1} f_n$ .
$$(g \circ f)_p : x \mapsto \sum_{ p \geq n \geq 1} \sum_{m \geq k_i \geq 1 \atop k_1 + ... + k_n = p}  g_n (f_{k_1} (x), ...,f_{k_n} (x) )$$
Let us state briefly these definitions.

\begin{defi}
Let us define $\oc _{1}E$ as the \lct $\bigoplus_{n \geq 1} \nmon (E, \KK)' $.
\end{defi}

\begin{defi}
For $ f \in \lin ( E_w,F_w)$ we define 
\begin{equation*}
\oc_1 f: 
\left\lbrace
\begin{split}
\oc_1 E_w & \rightarrow \oc_1 F_w \\
\phi & \mapsto ( (g_n) \in \prod_{n \geq 1} \nmon (F, \KK) \mapsto \phi ( (g_n \circ f)_n)
\end{split}
\right.
\end{equation*}
\end{defi}\medskip

\noindent This makes $\oc _{1}$ a functor on $\Weak$.
Such an exponential leads to a co-Kleisli category of non-unit sequences of monomials.
$$ \lin ( \oc _1 E_w, F_w)  \simeq \prod_{n \geq 1} \nmon ( E_w, F_w)_w .$$
It is endowed with the usual co-unit, and co-multiplication allowing for a composition which corresponds intuitively to a substitution.
 
\begin{prop}
The functor $ \oc _1: Lin \rightarrow Lin$ is a co-monad. Its co-unit $\epsilon: \oc _1 \rightarrow 1$ is defined by 
\begin{equation*}
\epsilon_E 
\left\lbrace
\begin{split}
\oc _1E_w & \rightarrow E_w \\
\phi  & \mapsto \phi_1 \in E'' \simeq E
\end{split}
\right.
\end{equation*}
The co-unit is the operator extracting from $\phi \in \oc _1E$ its part operating on linear maps.
The co-multiplication $\delta: \oc _1 \rightarrow \oc _1 \oc _1 $ is defined by 
\begin{equation*}
\delta_E 
\left\lbrace
\begin{split}
\oc _1 E_w \simeq (\prod_{p \geq 1} \pmon ( E, \KK ))' & \rightarrow \oc _1 \oc _1 E_w \simeq \left( \prod_{n \geq 1} \nmon ( [ \prod_{m \geq 1} \mmon(E, \KK)]', \KK) \right)' \\
\phi \in (\prod_p \pmon ( E, \KK ))'  & \mapsto (g_n)_n   \mapsto\\
&   \phi ( \left( x \in E \mapsto  \sum_{ p \geq n \geq 1} \sum_{m \geq k_i \geq 1 \atop k_1 + ... + k_n = p} g_n [ (f_m)_m  \mapsto g_n ( f_{k_1} (x), ...,f_{k_n} (x) ) ] \right)_p)
\end{split}
\right.
\end{equation*}
\end{prop}

The co-Kleisli category remains cartesian closed, and thus we obtain likewise a Seely isomorphism $$ \oc _1 (E_w \times F_w)  \simeq \oc _1 E_w \otimes \oc _{1} F_w .$$ As $\Weak$ is $\ast$-autonomous, we obtain this way another model of linear logic.

\section{Weak topologies and polarities}
\label{sec:polarities}

\subsubsection*{An interpretation for the shift connective.}
We chose to use weak topologies here in order to obtain a model of classical linear logic. Indeed, if $E$ is not endowed with its weak topologies, we have $E \sim E''$  but  this bijection is not bicontinuous.

As a consequence, we observe that negative and positive connectives of linear logic behave differently with respect to weak topologies. The interpretation of positive connectives must be endowed with their weak topology, whereas the interpretations of negative connectives are already endowed with their weak topology. Let us explain this idea.\smallskip

For negative connectives, we have: $E'_w \simeq E'$ (Proposition \ref{w_wstar}), $(E_w \times F_w)_w \simeq E _w \times F_w $ (Proposition \ref{weak_prod}), $ (E_w \parr F_w)_w \simeq E_w \parr F_w $ (Proposition \ref{cotensor}) and $ (\wn E)_w  \simeq \wn E$ (Proposition \ref{w_wstar}).

On the contrary, we do not have $(E_w \otimes F_w)_w \simeq (E_w \otimes F_w)$, nor $ (\oc E)_w \simeq \oc E  $. The case of the coproduct is more delicate, as when it is indexed by a finite set it corresponds with the product. We have $\bigoplus_{i \in I} (E_i)_w \simeq (\bigoplus_{i \in I} (E_i)_w)_w$ if and only if $I$ is finite (see Proposition \ref{weak_prod}).\smallskip

What we can conclude here is that we in fact constructed a model of the negative connectives of linear logic. Positive connectives are translated into negative connectives through a shift, and the shift is interpreted as the enforcement of the weak topology on some \lct. 

\begin{defi}
Formally, let us write $[|A|]$ the interpretation of a formula $A$ of $LL_{pol}$ as a \lct, and $[A]=[|A|]_w$ the interpretation of $A$ as an object of our model, that is a \lct endowed with its weak topology. 
\end{defi}

Let us recall the definition of $LL_{pol}$ \cite{Lau02}. 

\begin{center}
Negative formulas $N:= X^{\bot} ~|~  N \parr N ~|~ N \with N ~|~ \wn N ~|~ \shiftdown P $
\end{center}

\begin{center}
Positive formulas $ P:= X ~|~ P \otimes P ~|~ P \oplus P ~|~ \oc  P ~|~ \shiftup N $. 
\end{center}
The first interpretation $[| \cdot |]$ of a formula as an \lct is easy, as notations of linear logic are inspired by mathematics: $[| A \times B |] \simeq [|A|] \times [|B|]$, $[| A \otimes B |] \simeq [|A|] \otimes [|B|]$, etc. Things differ when we interpret them as object of our model.

From the previous explanations it follows that:

\begin{prop}
For negative formulas, the interpretation in our model is straightforward: when $N$ is a negative formula $[|N|]=[N]$.  
\end{prop}

\begin{prop}
When $P$ is a positive formula, then $[P]\simeq [\shiftdown P] \simeq [|P|]_w$. In general, we do not have $[|P|] \simeq [P]$, except for finite coproduct.     
\end{prop}

We have no mathematical interpretation of $\shiftup$, as the interpretations of negative formulas bear no other strong topology than their weak topology. All \lct can be seen as positive, but not all of them can be seen as negative, that is not all of them are endowed with their weak topology. Our model consists of negative \lct.
 
\subsubsection*{Double orthogonalities and polarities.} 

The interpretation of $A^{\bot}$ in our setting is $[A]'$, which is in no way an orthogonal of $A$. 
From this, it follows that $E \otimes F$ is not defined as as $(E \multimap F')'$, as its usual in denotational semantics  $(E \multimap F')'$  \cite{Ehr02,Ehr05,Bar00}, nor completed \cite{Gir96,BET12}, but as an algebraic tensor product endowed with some polarity. The fact that $E \otimes F$ is not constructed as the double-dual of some other \lct is responsible for the fact that we can see it as positive, that is $(E_w \otimes F_w) \nsimeq (E_w \otimes F_w)_w$. 

The same phenomenon happens for the exponential: we could have defined $ \oc  E $ as $ (\prod_n \nmon (E, \KK))'$ instead of $\oplus_n \nmon (E, \KK)'$, and this would have made it a negative connective, that is a connective which is already endowed with its weak topology.

\section*{Conclusion and further work}

We obtain a very general model of linear logic, using spaces which are commonly used in mathematics. We hope this work helps to initiate studies on computational interpretations of various theories used within the theory of topological vector spaces. The algebraic structure allows us to interpret the connectives of linear logic, whereas the topology on our spaces interprets the duality of classical logic and the polarities of the connectives. This paper appeals to further work on the relationships between weak spaces, polarised linear logic and focalized proofs \cite{And92}. The understanding of the shift as an operation of weak topology can also help to understand the decomposition of the exponential under polarities \cite[Section 4]{Lau04}.

As suggested by Barr's work \cite{Bar00}, we could try to construct a similar model of linear logic with Mackey spaces, that is spaces endowed with their Mackey topology. It would not interpret polarities, but could have other interesting properties. The key point in this construction would be to have an adaptation of Proposition \ref{fund_prop}. Of course, one could try and do this construction with any polar topology. Similarly, one can try to adapt the use of weak topologies to coherent Banach Spaces or convenient spaces to produce a model of linear logic. 

Another direction of research would be to construct a model with reflexive spaces as used in the literature, that is \lct which are isomorphic to their bidual when the dual is endowed with the topology of uniform convergence on bounded sets. These spaces are indeed endowed with a weak integral, and would allow for a step towards the understanding of the computational meaning of differential equations. One can note that a nuclear space which is also Fr\'echet or (DF) is reflexive, and the duality between Fr\'echet spaces and (DF)-spaces  nicely interprets the notion of polarity in linear logic. Constructing a model for the exponential of differential linear logic in the category of nuclear spaces is currently work in progress. 
 
 \subsection*{Acknowledgement}
 
I wish to thank T. Ehrhard for proof-checking this article, and for his suggestions on the exponentials. I am grateful to the referees for their many corrections and useful suggestions, particularly about the introduction of this paper.
\newpage

 
\bibliographystyle{alpha}
\bibliography{biblioWeakstar}

\end{document}